# Time delay in a disordered topological system

Yuhao Kang[1,2], Azriel Z. Genack[1,2*]

[1]Department of Physics, Queens College of the City University of New York, Flushing, NY 11367, USA

[2]Graduate Center of the City University of New York, New York, NY 10016, USA

Email: *genack@qc.edu

Abstract: The discovery of topological insulators has opened new prospects for robust signal transport for electronic, phononic, and photonic devices. Though transport of topological protected edge states is robust to disorder, large fluctuations and lengthened average delay time are observed. Here, we consider a quasi-1d system following the Haldane model and generalize the idea of eigenchannel time delay to the topological system. Eigenchannel time delay indicates the excited density of states for the configuration and relates to the intensity integral inside the system. Taking advantage of this property, we point out a practical way to extract the central frequency and linewidth of localized modes excited in the topological system. This work links the fluctuation of time delay to the strength of disorder and discusses the scaling of time delay.

A hallmark of topological insulator is the immunity of the edge mode to backscattering. Such immunity has been demonstrated numerically and experimentally for point or isolated defects [1–4], however the transport time of signals across a TI with randomly distributed disorder has not been explored in depth. Many applications of Tis depend not only on the robustness of transmission, but also on the wave propagation being ballistic with constant delay time and no extra modes excited in the medium.

Disorder creates quasi-normal modes inside the bulk region, which interact with the edge mode and govern the propagation of the edge mode. In a trivial system, the ratio of the average width and spacing in energy of modes, known as the Thouless number, $\delta=\delta\omega/\Delta\omega$, equals the dimensionless conductance or transmittance of the system. Here, we use $\omega$ to indicate the frequency in the photonic system or energy in the electronic system. When the configuration is strongly localized, the first transmission eigenchannel is strong correlated with resonant localized modes[5]. The quasi-normal mode[6–8] can be described by the complex energy $\omega - i\frac{\Gamma}{2}$, with central energy $\omega$ and linewidth $\Gamma$. When the sample is localized or in the crossover to localization, the spectrum of conductance or transmission can be approximated as the superposition of Lorentzian lines. The central frequency and linewidth of the quasi-normal modes can be estimated from these resonant peaks. The underlying quasi-normal modes can also be found from the modal decomposition of the complex field using the method of harmonic inversion[9].

However, in a TI system, the presence of robust edge mode erases the resonant peak of the transmission spectrum. This presents a challenge to the goal of uncovering the modes of the medium, which interact with the edge channel.

Here, we simulate propagation in a honeycomb lattice with Hamiltonian[10]

$$H = \sum_i c_i^\dagger \eta_i c_i + \sum_{\langle ij \rangle} c_i^\dagger t c_j + \sum_{\langle\langle ij \rangle\rangle} c_i^\dagger i\lambda_{SO} v_{ij} c_j$$

, where $c_i^\dagger (c_i)$ denotes the creation (annihilation) operation at the i$^{th}$ site, the onsite energy $\eta_i$ is uniformly distributed within –W/2 and W/2, t =1 is the hopping between nearest sites

$\langle ij \rangle$, $\lambda_{SO}$ is the spin-orbit coupling strength and $v_{ij} = (\hat{d_1} \times \hat{d_2})/|\hat{d_1} \times \hat{d_2}|$, where $d_1$ and $d_2$ are two nearest neighbor bonds connecting the next-nearest-neighbor $\langle\langle ij \rangle\rangle$. As shown in Fig. 1a, the scattering region is attached to two semi-infinite pristine leads. The upper/lower boundary supports a right/left-moving edge state. The signal is injected through the upper boundary. Simulations are carried out using an open-source package Kwant[11]. The lattice constant is 1.

Due to the finite transverse dimension of the system, excited modes in the medium can couple to the edge mode at either the upper or lower edge of the sample. The latter case causes a drop of transmitted intensity. A plot of a typical Green's function between the input and output at the upper boundary is shown in Fig. 1b. The magnitude is given by the blue curve and the phase by the red curve. The magnitude is a linear baseline with many dips. The wave amplitude in the interior for the three energy points in the flat region of the Green's function are similar, as shown in Fig. 1d-f. Line resonances are excited along the boundary so that the effective continuum state is the combination of the edge mode and these resonances. The resultant continuum state is not confined to the boundary. For example, in the circled region of Figs. 1e,f, a number of cavity modes are excited far from the upper boundary, over a large energy range. The case of a localized mode excited in the middle of sample is shown in Fig.1g. This cavity mode bridges the two boundary layers and cause the dip in the spectrum. Each dip corresponds to the excitation of a localize mode.

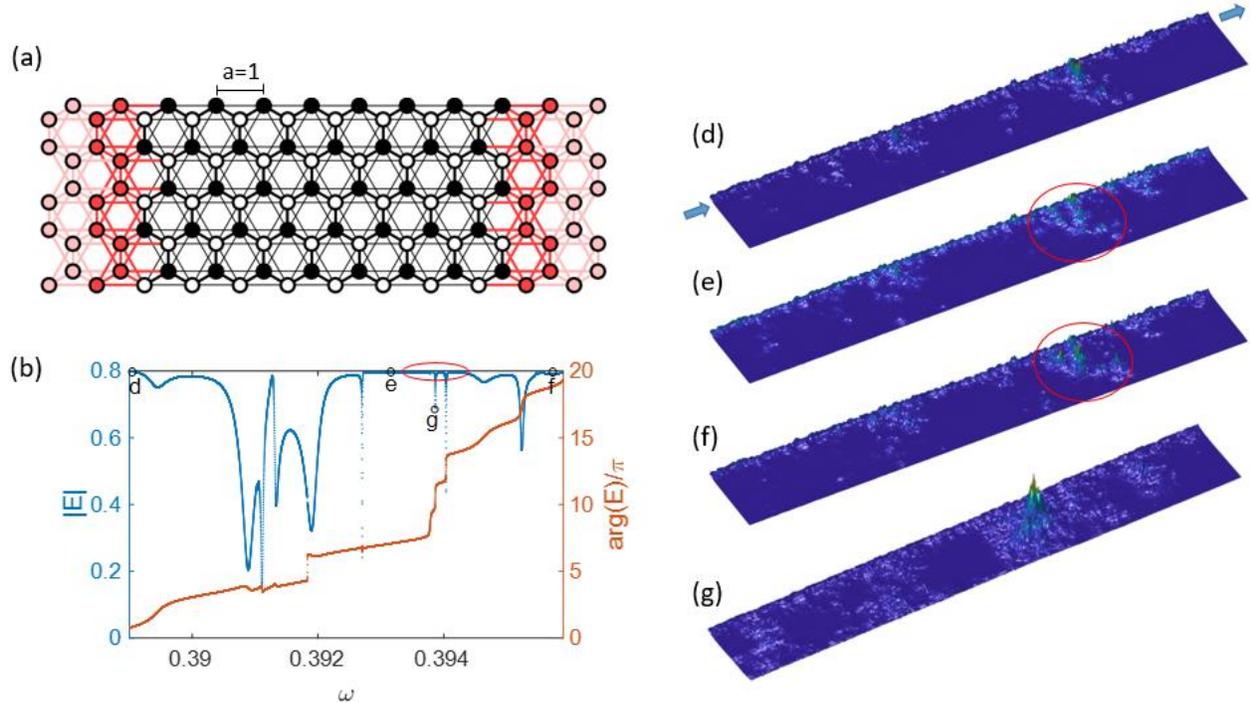

Fig. 1. Numerical simulation of disordered Haldane model. (a) Tight binding structure. The red region at the ends of the sample are two pristine leads. (b) The amplitude of the transmitted field at the output in the upper boundary. The blue curve gives the magnitude of the field and blue curve gives the corresponding phase. (d-g) The intensity distribution at four frequency points indicated in (b).

The transmitted field can be expressed by the superposition of quasi-normal modes[8]

$$E = E_0 + \sum_n \frac{V_n}{i(\omega_n - \omega) + \Gamma_n / 2},$$ where $E_0$ is the slowly-varied background due to the continuum state, $\omega_n$ and $\Gamma_n$ are the central frequency and linewidth of the $n^{th}$ discrete mode respectively. The second term is a resonant term, which describes the contribution of localized modes inside the bulk region.

Generally it is difficult to analysis the continuum term $E_0$. It is consist of two parts: the edge mode in the clean system; the disordered states created near the boundary, which form broad resonance due to the coupling to the edge mode. As a result, $E_0$ is a normalized edge term including the effect of disorder. Within a narrow energy range, this edge term can be approximated by linear fitting.

We focus on the spectrum within the red circle in Fig.1b and replot it in Fig. 2a. The circled line shows the transmitted signal. To estimate the contribution of the direct coupling term, we fit the magnitude and phase of $E_0$ and plot as the thin line in Fig. 2a. For spectrally isolated cavity modes, the phase increases by $2\pi$ as the frequency is tuned through a localized mode. In a random media, this phase change is $\pi$. Here, the abnormal phase change can be understood by the complex representation of field in the inset of Fig.2a. The edge term $E_0$ makes the curve of complex field encircle the original point. When the frequency is tuned though this circle, phase change would be $2\pi$. The edge mode is a slowly varied background. Over a broad frequency range relative to the discrete mode, the speckle pattern of edge mode in the interior of the sample does not change, so does the delay time of the edge mode, which is the integral of intensity inside the medium. As a result, aside from the region near the resonant frequency, the slope of the phase spectrum, which is the dwell time of the edge channel, is nearly constant, as can be seen in Fig.1b. To determine the phase of $E_0$, we first choose a region without the interruption of a cavity mode and fit a line to the phase variation. The magnitude of $E_0$ can be obtained through a linear interpolation of the baseline shown in Fig. 2a. Removing the contribution of the continuum term $E_0$, we plot the remaining resonant term in Fig.2b, and fit this spectrum using the method of harmonic inversion. Fig. 2c shows the contribution of each modes separately. The asymmetrical shape in Fig.2b indicates the Fano interference between the narrow and broad modes.

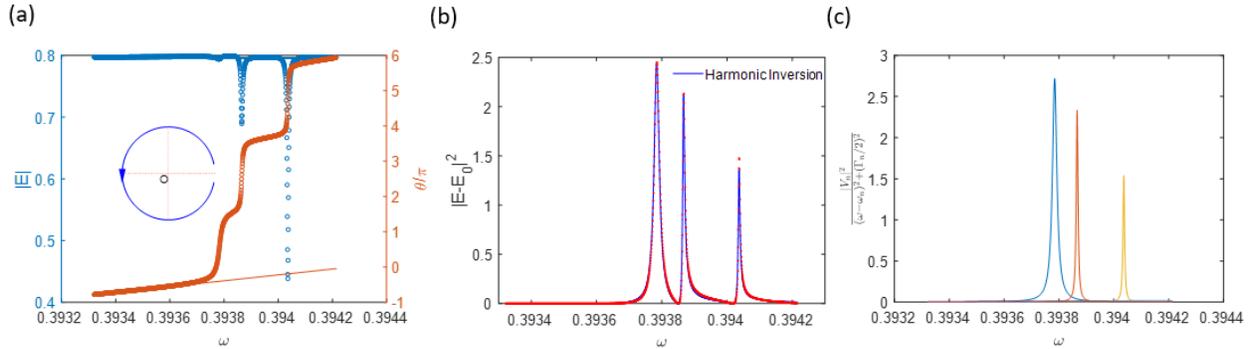

Fig. 2. Field decomposition. a) Zoom-in of region circled in red in Fig.1b. The curve made up of circles represents the transmitted field and the thin line means the fitted edge-mode term $E_0$. b) The red dotted line is the intensity of $E-E_0$, the blue line is determined by the fit to modes found using harmonic inversion. c) The contribution of each mode $|\frac{V_n}{i(\omega_n - \omega) + \Gamma_n/2}|^2$.

The applicable range of the field decomposition is limited. The distribution of energy density and the associated dwell time of continuum state change over a large frequency range, so that the estimation of phase of $E_0$ requires a changing background dwell time. The limitation of the spectral decomposition of the transmitted signal raises the problem of how to effectively extract the central frequency and linewidth of quasi-normal modes inside the system. To address this question, we consider the density of states (DOS) and delay time of the system. DOS can be calculated through Wigner-Smith matrix, $\rho = -\frac{i}{2\pi} Tr\left(S^\dagger \frac{dS}{d\omega}\right)$. The scattering matrix $S = \begin{bmatrix} r & t' \\ t & r' \end{bmatrix}$. Its elements are the field transmission coefficients between orthonormal incident and outgoing channels. The transmission matrix, $t$, gives the fullest account of transmission. The transmittance can be expressed as the sum of the transmission eigenvalues $\tau_n$ of the matrix product $T = Tr\, tt^\dagger = \sum_{n=1}^{N} \tau_n$.

The scattering matrix can be decomposed as[12]

$$S = \begin{pmatrix} W & 0 \\ 0 & U \end{pmatrix} \begin{pmatrix} -\sqrt{1-\tau} & \sqrt{\tau} \\ \sqrt{\tau} & \sqrt{1-\tau} \end{pmatrix} \begin{pmatrix} V^\dagger & 0 \\ 0 & X^\dagger \end{pmatrix}, \quad (1)$$

$\sqrt{\tau}$ is a diagonal matrix with elements equals the square root of $\tau_n$. $W, U, V, X$ are unitary matrices which indicate the mapping between eigenchannels and the orthogonal channels at the outgoing and incident sample interfaces. Generally, we need transmission matrix from both sides to calculate the total DOS. In Appendix A we show that $\rho = \frac{1}{2\pi} Im Tr(U^\dagger \frac{dU}{d\omega} - V^\dagger \frac{dV}{d\omega} + W^\dagger \frac{dW}{d\omega} - X^\dagger \frac{dX}{d\omega})$, where $U, V, W, X$ can be obtained through the singular value decomposition (SVD) of $t$ and $t'$, $t = U\sqrt{\tau}V^\dagger$ and $t' = W\sqrt{\tau}X^\dagger$.

When the system is reciprocal, DOS can be simplified to $\rho = \frac{1}{\pi} Im Tr\left(U^\dagger \frac{dU}{d\omega} - V^\dagger \frac{dV}{d\omega}\right)$. We use $u_n$ and $v_n$ to denote the $n^{th}$ column of matrices $U$ and $V$. The transmission time of the $n^{th}$ eigenchannel[13] is defined as $\frac{d\theta_n}{d\omega} = \frac{1}{i}(u_n^\dagger \frac{du_n}{d\omega} - v_n^\dagger \frac{dv_n}{d\omega})$. In a reciprocal system, the sum of $\frac{d\theta_n}{d\omega}$ for all eigenchannels is the DOS, $\rho = \frac{1}{\pi}\sum_n \frac{d\theta_n}{d\omega}$. $\frac{d\theta_n}{d\omega}$ is equal to $\pi$ times the contribution of the $n^{th}$ eigenchannel to the DOS. The transmission matrix is based on the orthogonal channels supported by the lead. We can also construct the matrix of retarded Green's function $G^R_{qp}(y_q, y_p)$ in real space, p/q indicates the incident and outgoing lead, $y_q$ and $y_p$ are the transverse coordinates at the two leads. $G_{qp}$ can be directly measured by recording the field transmission coefficients when the source and detector are scanned through all inputs and outputs. Fisher-Lee relation connects the Green's function matrix and transmission matrix, $t_{nm} = i\hbar\sqrt{v_n v_m}\iint \chi_n^\dagger(y_q)\left[G^R_{qp}(y_q, y_p)\right]\chi_m(y_p)dy_q dy_p$. $\chi_n$ and $v_n$ denote the normalized shape and group velocity of the $n^{th}$ channel in the lead. Similarly, we can calculated the SVD of the Green's function $G = U_G \Lambda_G V_G^\dagger$ and construct $\frac{d\theta_{Gn}}{d\omega} \equiv \frac{1}{i}(u_{Gn}^\dagger \frac{du_{Gn}}{d\omega} - v_{Gn}^\dagger \frac{dv_{Gn}}{d\omega})$, where $u_{Gn}$ and $v_{Gn}$ denote the $n^{th}$ column of matrices $U_G$ and $V_G$. Because $t$ and $G$ are not unitarily equivalent, for the $n^{th}$ eigenchannel $\frac{d\theta_{Gn}}{d\omega} \neq \frac{d\theta_n}{d\omega}$. However, in Appendix B we show that $Im Tr\left(U^\dagger \frac{dU}{d\omega} - V^\dagger \frac{dV}{d\omega}\right) = Im Tr\left(U_G^\dagger \frac{dU_G}{d\omega} - V_G^\dagger \frac{dV_G}{d\omega}\right)$. So the total DOS can be calculated from either method. When the system only support single channel in the lead, we can use Green's function $G$ to directly obtain the eigenchannel time delay.

The eigenchannel time delay is also related to the energy deposited inside the system. In the following discussion, we omit the subscript n since the system only supports a single input channel. We first consider a disordered trivial system attached to two single-channel leads in Fig. 3a. As shown in Fig.3b, eigenchannel time delay equals the average of the integral of the intensity for excitation from the left and right side, $\frac{d\theta}{d\omega} = \frac{1}{2}\int(I_{left} + I_{right})dA$. The inset in Fig.3b shows the transmission spectrum.

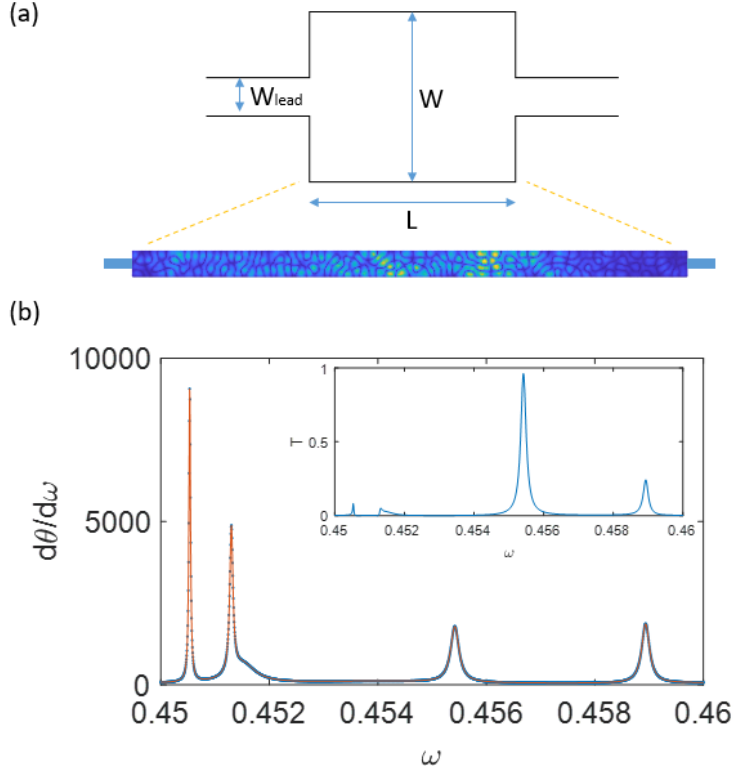

Fig. 3. Eigenchannel time delay in a random media system. a) The structure of a trivial system and a speckle pattern. $W_{lead}=6$ so that lead only supports single channel, W=20, L=400. b) The red curve is the eigenchannel time delay. The blue curve is the sum of intensity integral when the wave comes from both side divided by 2. The inset shows the transmission spectrum.

To generalize the relation between time delay and intensity integral to TI system, we consider two scenarios: 1) the edge modes at upper and lower boundary are not coupled by the disordered modes inside the bulk, 2) band gap is partially washed out due to strong disorder or finite transversal dimension of the sample.

In the first case, the upper boundary can be seemed as an isolated region with a left-incident channel and a right-outgoing channel. The transmission is near unity. Transmission matrix from left side becomes the whole scattering matrix for upper region. The DOS excited by the upper edge mode can be expressed by $\rho = \frac{1}{2\pi}\frac{d\theta}{d\omega}$. In a reciprocal system, this pre-factor should be $\frac{1}{\pi}$. We simulate a system with a weak disorder strength W=1.6. In Fig.4a, we compare $\frac{d\theta}{d\omega}$ with the integral of intensity excited from left side $\int I_{left} dA$ in the unit of $2\pi$.

When transmission is unity, these two quantities matches well. Around ω=0.3965, there is a small mismatch as shown in Fig.4b. We plot the speckle pattern at this energy in Fig.4c, at which a mode is excited near the lower boundary and the wave is scattered back at the lower edge of the sample via a localized state in the medium. This is further demonstrated in the spectrum of reflection time defined as $t_r = \frac{d\arg(r)}{d\omega}$. In Fig. 4d, the reflection time has a peak near this resonance. When the upper and lower boundary are decoupled, $\frac{d\theta}{d\omega}$ represents the DOS for the edge modes from one side as well as the excited bulk modes.

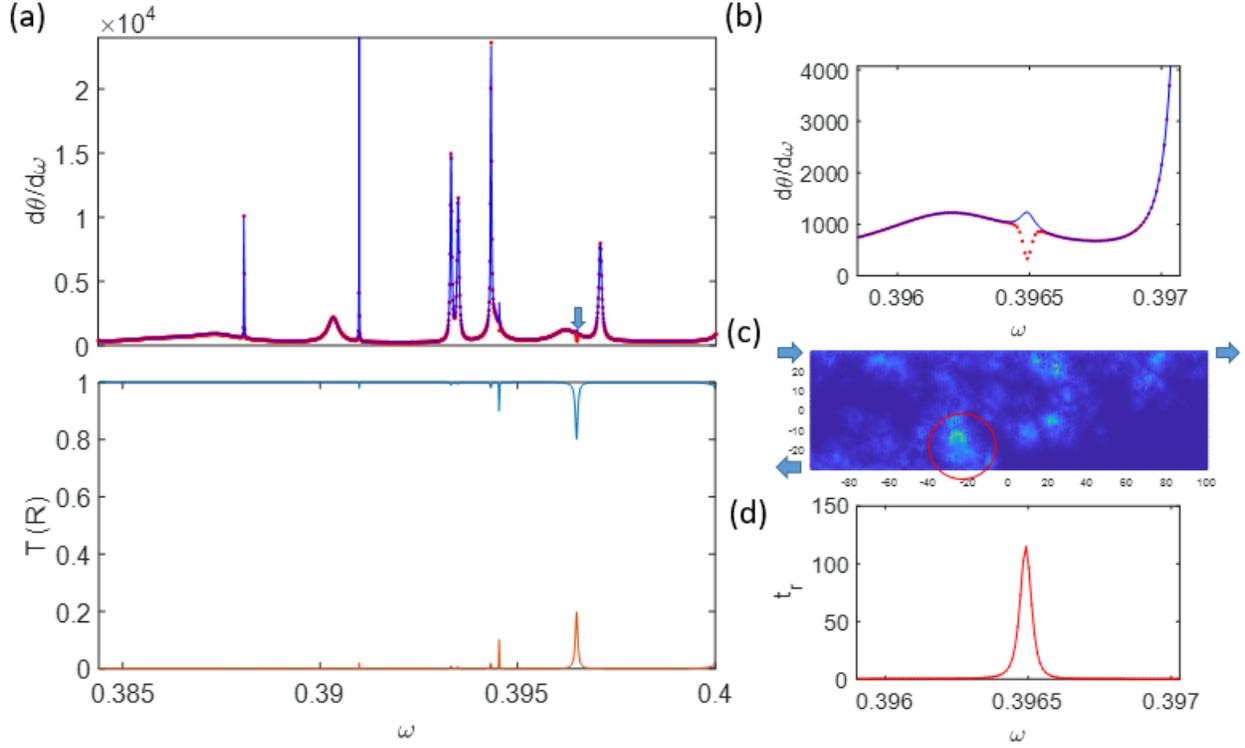

Fig. 4. Eigenchannel time delay in the Haldane model. (a) The red curve is the time delay and blue line is the intensity integral when the wave is injected form the left side. The transmission (blue) and reflection (red) spectra are shown in the lower panel. (b) Zoom-in of (a) in the region indicated by an arrow. (c) The intensity distribution at 0.3965. A mode in the red circle is excited and causes reflection. The corresponding reflection spectrum is shown in d).

The spectrum of DOS can be decomposed to the superposition of Lorentzian lines[14]

$$\frac{1}{2\pi}\frac{d\theta}{d\omega} = \rho_0(\omega) + \frac{1}{\pi}\sum_n \frac{\frac{\Gamma_n}{2}}{\left(\frac{\Gamma_n}{2}\right)^2 + (\omega - \omega_n)^2} \qquad (2)$$

The first term denotes the density of edge channel, which can be approximated as a constant term over a broad range. The second term describes the contribution of quasi-normal modes, which would be delta-functions in a closed system. In practice, there are also some modes that are outside the frequency range studies, their contribution would be small and vary slowly with frequency, which could be handled with a slowly varying polynomial term.

We use Eq. (2) to fit the time delay spectrum in Fig. 4a and display the decomposition result in Fig.5. The green curves shows the contribution of each of the mode. This fitting is based on an open-source peak-fitting program[15]. The peak in the time delay is associate with the resonance of the cavity mode.

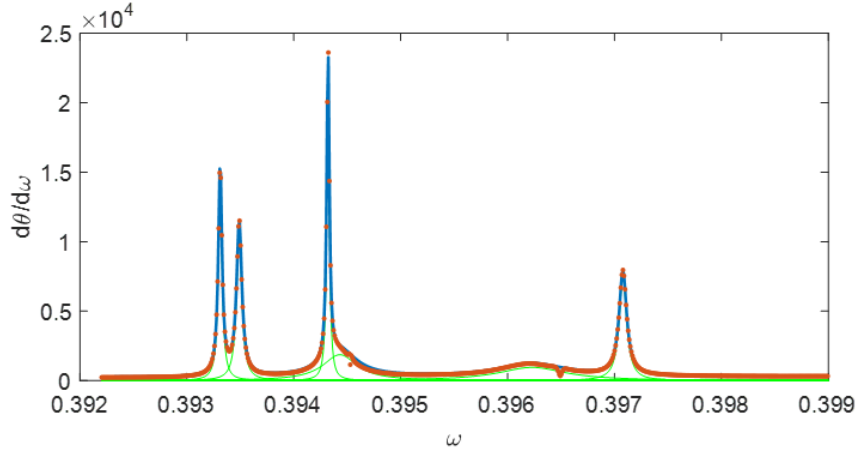

Fig.5. Lorentzian decomposition of the time delay. The red dotted curve is the time delay and the blue line is the fit of Eq. (2) to the time delay. The greens curves are the contribution of individual modes. There is a constant background around 200 due to edge channel and modes outside of this frequency range that is not shown.

The central frequency and linewidth of the modes can be retrieved from the spectral decomposition of the time delay. In this simulation, the peak is ~$10^4$ and the background is ~$10^2$. The edge term $\rho_0(\omega)$ in Eq.(2) can be ignored in the application and will not affect the fitting of the frequency and linewidth of mode. It can be seen that this method works for a broad frequency range and does not need the accurate estimation of the edge mode term.

When the reflection is not negligible, the wave coming in from the right need to be included. In Fig.6, we compare the sum of transmission time from both sides with the intensity integral from both sides, $\frac{d\theta}{d\omega} + \frac{d\theta'}{d\omega} = \int (I_{left} + I_{right}) dA$.

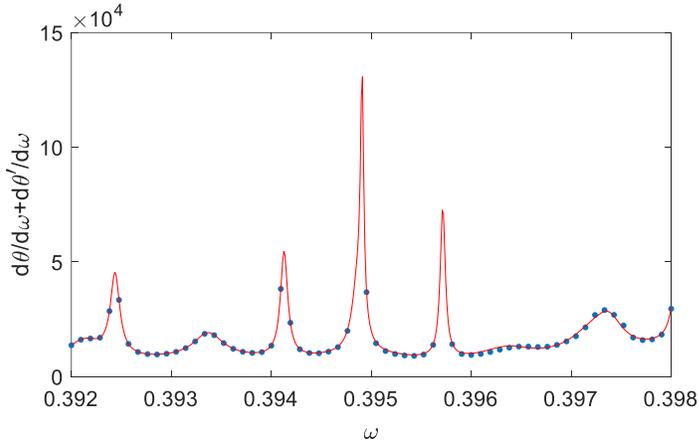

Fig. 6. The time delay spectrum in a strongly disordered configuration. The blue dotted curve is the sum of time delay from both sides and the red curve is the intensity integral when the wave comes from both sides.

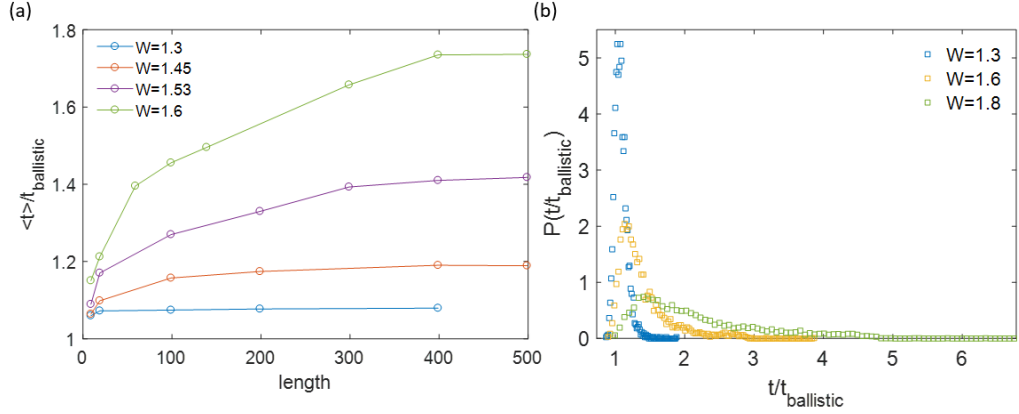

Fig. 7. Scaling and statistics of time delay. (a) The scaling of time delay relative to sample length and disorder strength. Width of system is 60. (b) Probability of density distribution of normalized time delay in three disordered cases. Length is fixed at 100.

In real applications, when the sample is wide or the disorder comes from fabrication defects, eigenchannel time delay is enough to represent the intensity integral inside. We compare the scaling and statistics of delay time in sample with different disorder strength. Time delay is normalized by the ballistic time in the clean sample. Fig. 7a shows the scaling of the average time delay for different strengths of disorder. When the sample is short, the delay time is close to the ballistic time. In this case, the configuration cannot support long lived modes in the cavity since the wave can couple out through the leads and so modes are broad. As the sample length increases, the normalized time delay increases and finally saturates. Time delay is proportional to the scattering region, which is determined by the decay length of the edge mode into the disordered region. Thus, this saturate ratio indicates the transverse expansion of wave. In Fig. 7b, we plot the probability density distribution of time delay for the three disorder strength at a length $L=100$. When the disorder is weak, the time delay is sharply peaked near the ballistic time. For stronger disorder, the distribution of time delay broadens and the average time delay is also increased.

In conclusion, we calculate the time delay of a disordered topological insulator confined as quasi-one-dimensional strip based on the tight-binding model. Two approaches, the field decomposition and time delay decomposition, are tested to extract the complex energy of the discrete modes in the presence of the edge mode. The previous discussion of time delay and DOS are mainly based on the scattering matrix, we show here that the spatial matrix of Green's function between the output and input can be directly used to calculate the DOS. This facilitates the calculation of DOS because it is easier to obtain the Green's function, in measurements and numerical simulation such as recursive Green's function method. When modal overlap is small, the transmitted field can be decomposed into a superposition of discrete modes and the edge channel. This decomposition is possible even though transmission is dominated by robust transmission in the edge state. When the upper edge mode is not coupled to the lower edge mode via disorder, the eigenchannel time delay is then equivalent to the integral of intensity inside the sample, whose spectrum can be decomposed into a superposition of Lorentzian lines. The saturate ratio between the average eigenchannel time delay and the ballistic time in a long sample provides an index to describe the average spread of the edge mode in disordered case. The modes inside the bulk region determine the dynamics of signal transport. This work provides a way to analyze the bulk modes from the transmitted signal and can be generalized to the time-invariant TI system.

## Appendix A. Calculation of DOS based on transmission matrix

The density of states can be calculated through the Wigner-Smith matrix

$$\rho = -\frac{i}{2\pi}Tr\left(S^\dagger \frac{dS}{d\omega}\right) = \frac{1}{2\pi}ImTr(t^\dagger \frac{dt}{d\omega} + r^\dagger \frac{dr}{d\omega} + t'^\dagger \frac{dt'}{d\omega} + r'^\dagger \frac{dr'}{d\omega})$$

(A1)

Following the notation in Eq. (1), transmission matrix $t = U\sqrt{\tau}V^\dagger$

$$ImTr\left(t^\dagger \frac{dt}{d\omega}\right) = ImTr\left[V\sqrt{\tau}U^\dagger\left(\frac{dU}{d\omega}\sqrt{\tau}V^\dagger + U\frac{d\sqrt{\tau}}{d\omega}V^\dagger + U\sqrt{\tau}\frac{dV^\dagger}{d\omega}\right)\right]$$

$$= ImTr\left(\tau U^\dagger \frac{dU}{d\omega} + \sqrt{\tau}\frac{d\sqrt{\tau}}{d\omega} + \tau\frac{dV^\dagger}{d\omega}V\right)$$

$$= ImTr[\tau\left(U^\dagger \frac{dU}{d\omega} - V^\dagger \frac{dV}{d\omega}\right)]$$

(A2)

Similarly, we can write out the other three terms and get a general expression for DOS

$$\rho = \frac{1}{2\pi} ImTr(U^\dagger \frac{dU}{d\omega} - V^\dagger \frac{dV}{d\omega} + W^\dagger \frac{dW}{d\omega} - X^\dagger \frac{dX}{d\omega})$$

(A3)

where $t = U\sqrt{\tau}V^\dagger$ and $t' = W\sqrt{\tau}X^\dagger$.

When the system is reciprocal, $t^T = t'$, then $W^T = V^\dagger$ and $U^T = X^\dagger$, so that $Tr\left(U^\dagger \frac{dU}{d\omega}\right) = -Tr\left(U^T \frac{dU^*}{d\omega}\right) = -Tr(X^\dagger \frac{dX}{d\omega})$.

Similarly, $Tr\left(V^\dagger \frac{dV}{d\omega}\right) = -Tr\left(W^\dagger \frac{dW}{d\omega}\right)$. Thus the DOS for a reciprocal system can be simplified to

$$\rho = \frac{1}{\pi} ImTr\left(U^\dagger \frac{dU}{d\omega} - V^\dagger \frac{dV}{d\omega}\right)$$

$$= \frac{1}{\pi}\sum_n (-i)\left(u_n^\dagger \frac{du_n}{d\omega} - v_n^\dagger \frac{dv_n}{d\omega}\right)$$

(A4)

where $u_n$ and $v_n$ are $n^{th}$ column of $U$ and $V$.

Appendix B. Equivalence of transmission matrix and Green's function in terms of calculation of the DOS

We first rewrite the Fisher-lee relation in the matrix form, transmission matrix

$$t = i\hbar D_o \chi^\dagger G \chi D_i, \tag{B1}$$

We will take $\hbar$ to equal 1, in which case the results also fully apply to classical waves. $i$ will be omitted since we only consider $t^\dagger \frac{dt}{d\omega}$. $D = diag(\sqrt{v_1}, \sqrt{v_2}, \cdots \sqrt{v_n})$ is the velocity matrix of channels supported by lead. $D_o/D_i$ represents the velocity matrix at output/input. We carry out the singular value decomposition for both Green's function and transmission matrix,

$$G = U_G \Lambda_G V_G^\dagger, \quad t = U_t \Lambda_t V_t^\dagger, \quad (B2)$$

Construct $T_1$ and $T_2$ so that $U_t = D_o \chi^\dagger U_G T_1$ and $V_t^\dagger = T_2^\dagger V_G^\dagger \chi D_i$. Combined with Eq.(1) and Eq.(2), this gives

$$\Lambda_G = T_1 \Lambda_t T_2^\dagger. \quad (B3)$$

$U_t U_t^\dagger = I$ gives

$$D_o \chi^\dagger U_G T_1 T_1^\dagger U_G^\dagger \chi D_o = I. \quad (B4)$$

So that

$$\begin{aligned}
&\operatorname{Im} Tr(U_t^\dagger \frac{dU_t}{d\omega}) \\
&= \operatorname{Im} Tr[T_1^\dagger U_G^\dagger \chi D_o (\frac{dD_o}{d\omega} \chi^\dagger U_G T_1 + D_o \frac{d\chi^\dagger}{d\omega} U_G T_1 + D_o \chi^\dagger \frac{dU_G}{d\omega} T_1 + D_o \chi^\dagger U_G \frac{dT_1}{d\omega})] \\
&= \operatorname{Im} Tr[D_o^{-1} \frac{dD_o}{d\omega} + \chi^{\dagger -1} \frac{d\chi^\dagger}{d\omega} + U_G^{-1} \frac{dU_G}{d\omega} + T_1^{-1} \frac{dT_1}{d\omega}] \\
&= \operatorname{Im} Tr[\chi^{\dagger -1} \frac{d\chi^\dagger}{d\omega} + U_G^{-1} \frac{dU_G}{d\omega} + T_1^{-1} \frac{dT_1}{d\omega}]
\end{aligned} \quad (B5)$$

In the second line, we make use of $Tr(AB) = Tr(BA)$ and Eq. (4). In the third line, the first term is discarded because $D_o$ is a real matrix. $T_1^{-1}$ is left inverse of $T_1$. $\chi^{\dagger -1}$ is right inverse of $\chi^\dagger$.

Similarly,

$$\operatorname{Im} Tr(V_t^\dagger \frac{dV_t}{d\omega}) = \operatorname{Im} Tr[\chi^{\dagger -1} \frac{d\chi^\dagger}{d\omega} + V_G^{-1} \frac{dV_G}{d\omega} + T_2^{-1} \frac{dT_2}{d\omega}] \quad (B6)$$

Because $\Lambda_G$ is real and using Eq.(B3) gives

$$0 = \text{Im}\,Tr(\frac{d\Lambda_G}{d\omega}\Lambda_G^{-1})$$

$$= \text{Im}\,Tr(\frac{dT_1}{d\omega}\Lambda_t T_2^{\dagger}\Lambda_G^{-1} + T_1\frac{d\Lambda_t}{d\omega}T_2^{\dagger}\Lambda_G^{-1} + T_1\Lambda_t\frac{dT_2^{\dagger}}{d\omega}\Lambda_G^{-1})$$

$$= \text{Im}\,Tr(\frac{dT_1}{d\omega}T_1^{-1} + T_1\frac{d\Lambda_t}{d\omega}\Lambda_t^{-1}T_1^{-1} + T_1\Lambda_t\frac{dT_2^{\dagger}}{d\omega}T_2^{\dagger-1}\Lambda_t^{-1}T_1^{-1}) \quad (B7)$$

$$= \text{Im}\,Tr(\frac{dT_1}{d\omega}T_1^{-1} + \frac{d\Lambda_t}{d\omega}\Lambda_t^{-1} + \frac{dT_2^{\dagger}}{d\omega}T_2^{\dagger-1})$$

$$= \text{Im}\,Tr(T_1^{-1}\frac{dT_1}{d\omega} - \frac{dT_2^T}{d\omega}T_2^{T-1}) = \text{Im}\,Tr(T_1^{-1}\frac{dT_1}{d\omega} - T_2^{-1}\frac{dT_2}{d\omega})$$

Equations (B5-B7) give the final result

$$\text{Im}\,Tr(U_t^{\dagger}\frac{dU_t}{d\omega} - V_t^{\dagger}\frac{dV_t}{d\omega}) = \text{Im}\,Tr(U_G^{\dagger}\frac{dU_G}{d\omega} - V_G^{\dagger}\frac{dV_G}{d\omega})$$